\documentstyle[aps,prc,psfig]{revtex}
\begin{document}
\twocolumn[
\columnwidth\textwidth\csname@twocolumnfalse\endcsname
\title{Shape coexistence and tilted-axis rotation in neutron-rich hafnium isotopes}
\author{Makito Oi$^{1,2}$\footnote{m.oi@surrey.ac.uk}, Philip M. Walker$^{1}$, and Ahmad Ansari$^{3}$}
\address{$^{1}$ Department of Physics, University of Surrey, 
Guildford, Surrey GU2 7XH, United Kingdom. \\ 
$^{2}$Department of Applied Physics, Fukui University, 3-9-1 Bunkyo, 
Fukui 910-8507, Japan.\\
$^{3}$Institute of Physics, Bhubaneswar 751 005, India.}
\date{\today}
\maketitle 
\begin{abstract}
We have performed  tilted-axis-cranked Hartree-Fock-Bogoliubov
calculations for a neutron-rich hafnium isotope ($^{182}$Hf) whose
proton and neutron numbers are both in the upper shell region.
We study  whether the shell effects play a role in producing
high-$K$ isomers or highly gamma-deformed states at high spin.
In particular, the possibility of shape coexistence and
the effect of wobbling motion are discussed.
\end{abstract}

]

Hafnium isotopes ($Z=72$) are best known as nuclei that have 
high-$K$ isomers (e.g., the $K^{\pi}=16^+$ isomer
in $^{178}$Hf, with a half life
$t_{1/2}=31$ yr \cite{TOI96}). 
From a viewpoint of the Nilsson model,
a reason is that proton single-particle levels are filled
up to the upper part of the shell where there are 
many high-$\Omega$ states at prolate deformation
($\Omega$ is the angular momentum projection
on the nuclear symmetry axis). 
The presence of long-lived high-$K$ isomers 
indicates the existence of  axial symmetry to make $K=\sum_i \Omega_i$, 
a good quantum number.
With recent developments of experimental techniques, 
such as fragmentation \cite{PRP00} and 
deep-inelastic reactions \cite{WDC98} in populating high-spin states, 
the study of nuclei in the well deformed rare-earth region 
moves away from the $\beta$ stable line
towards the neutron-rich isotopes.
D'Alarcao et al. recently discovered several high-$K$ isomers 
in the neutron-rich hafnium isotope of $^{182}$Hf \cite{ACS99}.
For neutron-rich hafnium isotopes,
whose neutron Fermi level is located in a similar position 
to the proton one (i.e., the upper half of the shell), 
 we can  expect an even more important role for
 high-$K$ isomeric states \cite{WD99}.
At the same time, however, 
the empty nucleon states near the Fermi surfaces
can be considered as hole states, and these
 may induce substantial gamma deformation 
which breaks the axial symmetry.

A few theoretical investigations of high angular momentum
collective states have been reported for such neutron-rich hafnium
isotopes:
by means of a  microscopic method 
(cranked Hartree-Fock-Bogoliubov method, or C-HFB),
the possibility of a collective oblate deformation 
at high-spin ($I=26\hbar$) in $^{180}$Hf was predicted \cite{HM79};
by means of a macroscopic-microscopic method 
(total routhian surface calculations, or TRS),
the existence of non-collective prolate-deformed states
 becoming yrast, i.e., lowest in energy, for $I \agt 10\hbar$ 
was shown  in $^{182-186}$Hf \cite{XWW00}. 
In the latter study, the collective
oblate deformed states become  yrast 
when quite high in spin ($I=36\hbar$ in $^{182}$Hf).
In both of the studies, the rotational states are treated only by the
one-dimensional cranking model, but
recent developments in the tilted-axis-cranking model \cite{Ons86,Fr93,MMM00} 
allow for the possibility of further investigation of these
neutron-rich hafnium isotopes.

In this paper, 
in a microscopic framework of  the
two-dimensional tilted-axis-cranked HFB method,
we analyze the high-spin structures 
predicted for the neutron-rich hafnium isotope $^{182}$Hf
(which has $N/Z = 1.53$).

First, the calculational procedure is briefly described.
We solve the HFB equations with a 2d-cranking term self-consistently 
for $^{182}$Hf, following the method of steepest-descent \cite{Ons86}.
Nucleon numbers and the total angular momentum are constrained 
during the iterations;
$\langle \hat{N}_{\tau}\rangle = N_{\tau} \quad (\tau = {\rm p, n});
J_x = \langle \hat{I}_x \rangle = J\cos\theta;
J_y = \langle \hat{I}_y \rangle = 0;
J_z = \langle \hat{I}_z \rangle = J\sin\theta$.
(Note that the tilt angle $\theta$ is measured 
from the $x$-axis to see the
deviation of the rotation axis from the $x$-axis. 
In this study, it corresponds to
a principal axis of the quadrupole deformation, perpendicular
to the symmetry axis that is chosen to be the $z$-axis).
We also constrain off-diagonal components of the quadrupole tensor 
such that the intrinsic coordinate
axes coincide with the principal axes of the quadrupole deformation.
See Ref.\cite{Ons86}, for details.

The one-dimensional cranking calculations 
(or 1d-cranking calculations),
in which the rotation axis is fixed to be 
along the $x$-axis ($\theta=0^{\circ}$), are performed as follows.
First for $J=0$, 
the Nilsson + BCS state is taken as a trial state.
Then  cranked HFB states, $\Psi_{\rm u} (J)$, are calculated
up to $J=40\hbar$ with increments $\Delta J = 0.1\hbar$.
This way of calculation is what we call ``up-cranking'' calculations.
Then, using the up-cranking  solution at $J=40\hbar$ as a trial state,
``down-cranking'' calculations
are similarly performed from $J=40\hbar$ down to $J=0$ 
to obtain $\Psi_{\rm d} (J)$.

The calculations of tilted-axis-cranked HFB states (2d-cranked states,
or $\Psi^{t}(J,\theta)$) are carried out
by starting at a state $\Psi^{\rm t}(J,\theta=0^{\circ})$,
which are calculated through an up-cranking calculation,
and performing a ``forward'' tilting calculation up to $\theta=90^{\circ}$
with increments $\Delta\theta = 0.5^{\circ}$ for each (integer) $J$.
Then in a similar manner we make ``backward'' tilting calculations
from $\theta=90^{\circ}$  to $\theta=0^{\circ}$.
At certain tilt angles, the forward and backward tilting
results do show interesting differences, particularly, at high spins.

Our Hamiltonian consists of a spherical part ($H_0$) 
and a residual part ($V_{\rm res}$) 
employing the pairing-plus-$\rm Q\cdot Q$ interaction.
In the hafnium isotopes the hexadecapole interaction can be
important, but we have checked that our results below are not
affected very much by including the interaction. 
Thus we omit it in the present study.
The Hamiltonian is thus written,
\begin{eqnarray}
 H &=& H_0 + V_{\rm res}\\ \nonumber
   &=& \sum_{\tau={\rm p, n}}\sum_{i=1}^{N(\tau)}\epsilon_i c_i^{\dag}c_i
  -\frac{1}{2}\kappa\sum_{\mu=-2}^{2}\hat{Q}_{\mu}^{\dag}\hat{Q}_{\mu}
  -\sum_{\tau={\rm p,n}} G_{\tau}\hat{P}_{\tau}^{\dag}\hat{P}_{\tau},
\end{eqnarray}
in which $\epsilon_i$ means a spherical Nilsson level
and $i$ runs all over the model space. 
Force parameters $\kappa$ and $G_{\tau}$ are determined 
in the framework of the Nilsson + BCS model by giving input parameters
for the quadrupole deformation 
($\beta^{\rm ini}, \gamma^{\rm ini})$
and gap energy $(\Delta^{\rm ini}_{\tau}$).
(Our definition of $\gamma$ is taken from pp.8 in Ref.\cite{RS80},
and is opposite to the Lund convention in the sign.)
In this paper, we employ a set of the input parameters 
based on the calculations by M\"oller et al.\cite{MN95}
because we find it gives a good agreement with experimental data.
For $^{182}$Hf, we employ 
$\Delta^{\rm ini}_{\rm p}=0.725$ MeV;
$\Delta^{\rm ini}_{\rm n}=0.625$ MeV;
$\gamma^{\rm ini}=0^{\circ}$;
$\beta^{\rm ini} = 0.270$ or $0.268$.
We will see soon the reason for the two values of $\beta^{\rm ini}$.

Our single-particle model space is almost the same as 
the choice of Kumar and Baranger \cite{KB68}
(two major shells in the spherical Nilsson model: 
$N=4,5$ for proton and $N=5,6$ for neutron),
with two extra single-particle orbits,
proton i$_{13/2}$ and neutron $j_{15/2}$.
The single particle energies are the spherical Nilsson
model energies with $A$-dependent Nilsson parameters \cite{NPA131}.

First of all, let us discuss a principal axis rotation
($\theta=0^{\circ}$) in 1d- and 2d-cranking calculations.
Fig.1 shows the energy spectra of one-dimensional up- and down-cranking
states ( $ \Psi_{\rm u}(J) $ and $ \Psi_{\rm d}(J) $, 
respectively ) and the states of $\theta=0^{\circ}$ 
in the 2d-cranked calculations,$ \Psi^{t}(J,\theta=0^{\circ}) $,that
are obtained through the backward tilting procedure at a given
value of $J$.
They are plotted also in the inset of Fig.1
 together with the known experimental values of the g-band.
The g-band is well reproduced by our one-dimensional cranking calculations.

We have two kinds of 1d-cranked states plotted in Fig.1.
They are calculated with the same conditions 
except for the initial value of $\beta^{\rm ini}$.
One (denoted as ``A'' in the graph) is obtained 
with the initial value of $\beta^{\rm ini}=0.270$
while  the other (denoted as ``B'') is with $\beta^{\rm ini}=0.268$.
The small difference in $\beta^{\rm ini}$  gives 
almost no difference in the wave functions at low spin $(J\alt 16\hbar$),
 but it does lead to a significant difference at
 high spins.
The sensitive dependence of the high-spin HFB solutions 
to $\beta^{\rm ini}$ shows that the energy manifold in the
variational space near the crossing regions
has several local minima that are almost degenerate.
The states ``A'' and ``B'' imply the possibility of band crossings
\footnote{As Hamamoto et al. pointed out \cite{HN90},
the validity of the cranking model is questionable in the band
crossing region because of the
semi-classical aspects in the model. 
However, the model works well outside the
crossing region and could explain qualitative features near the
crossing region, such as alignments.
For a more acculate analysis of the crossing region,
we should employ the so-called ``variation-after-projection'' 
method \cite{RS80}, the generator coordinate method \cite{OAHO00},
or the diabatic method \cite{Ha77}.
}, 
which correspond to the regions $17\hbar \alt  J \alt 26\hbar$ and
at $22\hbar \alt J \alt 31\hbar$ respectively 
\footnote{Note that we do not mean here that these values 
of the crossing angular momentum correspond exactly to the experimental values.
In general, the simple self-consistent cranking calculations 
do not reproduce the value precisely\cite{Ma75}.}
in the calculations, but 
a substantial difference between the states is seen in the gamma deformation.
Before the crossing regions
both of the states have $\gamma\simeq 10^{\circ}$, 
while after the crossing regions
the states ``A'' have near-prolate deformation
with a  negative gamma value ($\gamma\simeq -10^{\circ}$) and
the states ``B'' have oblate shape ($\gamma\simeq 60^{\circ}$).
(See Fig.2(a) for evolutions in gamma deformation for each solution).
From an analysis of our numerical results, 
the oblate deformation is caused by the gradual alignments 
of both i$_{13/2}$ neutrons and h$_{11/2}$ protons,
while, in addition to these alignments, the near-prolate deformation 
with $\gamma\simeq -10^{\circ}$ is caused by the quick alignment of 
j$_{15/2}$ neutrons.
(Note that the neutrons in the j$_{15/2}$ orbits 
are not part of the usual Kumar-Baranger model space).
The solution ``A'' is reported in this paper for the first time.
Xu et al. found that there is no stable minimum corresponding to 
this solution in their TRS calculations \cite{XWW00}.
However, they performed their (1d) cranking calculations for
given rotational frequencies ($\omega _x$), while we have performed
calculations for given (average) angular momentum vectors ($J_{x},J_{y},J_{z}$).

In a spin region $23\hbar \le J \le 35\hbar$,
the three kinds of states having different gamma deformations
($\gamma\simeq \pm 10^{\circ}$ and $60^{\circ}$) 
are close to each other in energy.
This result implies  
a manifestation of shape coexistence at high spin,
or
 multi-band crossings among bands
specified by different gamma deformations (or corresponding
rotational alignments).

It is interesting to see the energies for the state 
$\Psi^{t}(J,\theta=0^{\circ})$,
 which are represented  by the thick solid line in Fig.1.
The line has two discontinuities,
 at $J=27\hbar$ and $34\hbar$, implying 
two configuration changes.
At lower spin ($J\alt 26\hbar$), 
$ \Psi^{t}(J,\theta=0^{\circ}) $
follows the 1d-cranking calculations which give rise to 
nearly prolate shape ($\gamma \simeq 10^{\circ}$ at $J\simeq 25\hbar$).
Then at $J=27\hbar$, the gamma deformation changes to 
$\gamma\simeq -10^{\circ}$, which is the same as ``A''.
Finally, at $J=34\hbar$, 
$ \Psi^{t}(J,\theta=0^{\circ}) $
changes into the same state as ``B'', having oblate shape
($\gamma\simeq 60^{\circ}$).
The result that the  states having different gamma deformation
are connected by the tilted-axis-cranking solutions 
indicates the importance of the tilting degree of freedom
($\theta$) for the search for excited states near band crossings.
In Fig.2(b), we show how the quadrupole deformations 
($\beta$ and $\gamma$) evolve at $J=34\hbar$ as we vary
the tilt angle ($\theta$).
The corresponding energy curves are plotted in Fig.3(a).
There are three types of solutions with
different gamma deformation:
(i)the solution having no $\theta$-dependence in tilt angles for
 $\theta\alt 20^{\circ}$ is oblate ($\gamma\simeq 60^{\circ}$);
(ii)the solution which shows 
a tilted rotation minimum at $\theta\simeq 10^{\circ}$ 
has negative gamma deformation ($\gamma\simeq -10^{\circ}$);
(iii)the solution with a minimum at $\theta=90^{\circ}$,
which may correspond to high-$K$ states, has $0^{\circ}\alt 
\gamma \alt 10^{\circ}$.

The energy difference between the states of type (i) and (ii) 
is roughly constant and small ($\simeq 500$ keV), 
so that these two states can couple to form states with mixed
deformation. 
The energy curves for these states
are shallow in the range $0^{\circ} \le \theta \le 30^{\circ}$,
so that fluctuations in the rotation axis, or 
 wobbling motion \cite{OAHO00}, can be expected.
However, gamma deformations for each state of type (i) and (ii)
are quite constant against variation in the tilt angle up 
$\theta\simeq 30^{\circ}$.
Therefore, rather than a picture in which  states of type (i) and
(ii) are coupled thorough the wobbling motion,
we should have a picture where
they are connected possibly through  $\gamma$
tunnelling, and where the mixed states 
wobble around the tilted rotation minimum at 
$\theta\simeq 10^{\circ}$.
Nevertheless, a coupling of these mixed states with states of type (iii),
is possible through wobbling motion.

It is interesting to see in Fig. 3(a) the energy curve at $J=40\hbar$
corresponding to the states of type (i) above.
There are two minima at $\theta=0^{\circ}$ and $90^{\circ}$, but
the barrier height between them is only $\simeq 50$ keV,
not visible in the plot.
The corresponding deformation is collective oblate (the 
$y$-axis is the symmetry axis) and 
almost constant over the entire range of $\theta$.
For a strict oblate symmetry, there would be no energy dependence
on $\theta$.
The projection of the nuclear deformation onto the $x-z$ plane
is a circle, so that there is no preference for a direction 
of (collective) rotation in the $x-z$ plane.
For finite triaxiality, it is important to consider wobbling motion
in the $\theta$ direction.

From these discussions, 
we can deduce that
the shape coexistence  creates  successive backbends
(sudden changes in moments of inertia) in the excited rotational bands
of $\theta\simeq 0^{\circ}$.
The first backbend, which is caused by neutron
j$_{15/2}$ alignment, in addition to alignments 
of neutron i$_{13/2}$ and proton h$_{11/2}$,
 is expected  at lower spins, 
which corresponds to $17\hbar \alt J \alt 26\hbar$
in our calculations.
The second one, which is
caused by de-alignment of the j$_{15/2}$ neutrons and 
retaining alignments of the i$_{13/2}$ neutrons and h$_{11/2}$ protons,
can be seen at higher spins (corresponding to $22\hbar \alt J \alt 31\hbar$
in the calculations).
According to Fig.1, 
the second backbend can be more pronounced than the first one.
It is also possible that these two backbends are mixed together
to create one giant backbend as Hilton and Mang predicted \cite{HM79},
but according to our analysis 
three types of gamma deformations are involved: 
$\gamma\simeq \pm 10^{\circ}$ and $\simeq 60^{\circ}$.

Now, let us look at the calculated high-$K$ states.
In Fig.1, states of $\theta=90^{\circ}$, 
$ \Psi^{\rm t}(J,\theta=90^{\circ}) $, are
shown with ``$+$'' symbols.
These states correspond to a local minimum at $\theta=90^{\circ}$ 
in the energy curve (see Fig.3(b)). 
In our calculations, 
this minimum starts to appear at $J\simeq 8\hbar$, and
becomes the lowest minimum at $J=12\hbar$ and at higher spins.
The $\theta=90^{\circ}$ minimum may be considered
approximately to correspond to a high-$K$ state,
and the corresponding $z$-axis cranked state, or 
$ \Psi^{\rm t}(J,\theta=90^{\circ}) $, 
is a simulation of 
nuclear  rotation where single-particle angular momenta carry
most of the total angular momentum.
We have checked that the corresponding gamma deformation is almost zero
as in Fig.2(b) which shows that there is axial symmetry for these states.

However, we should note that  $J_z (= J\sin\theta)$ 
in the self-consistent cranking calculations
is not an eigenvalue but just an expectation value, due to the
rotational symmetry breaking by the mean field.
Besides, in the tilted-axis-cranked HFB calculations,
angular momenta are fully mixed in the sense that
even a mixture among even and odd angular momenta happens.
This is because the signature symmetry, 
a discrete subgroup ($D_2$) of the rotational group $O(3)$,
is broken by tilted rotation \cite{OOTH98}.
We should therefore keep in mind that the mean-field description
of high-$K$ states has a certain limitation.

Remembering the above remarks, let us look at our results for 
$ \Psi^{\rm t}(J,\theta=90^{\circ}) $.
The experimental energy of the isomer, which is tentatively
assigned to $K^{\pi}=13^+$,
is 2.572 MeV relative to the ground state energy,
while the numerical values
are  2.101 MeV ($J=12\hbar$), 2.255 MeV ($J=13\hbar$), 
and 2.408 MeV ($J=14\hbar$).
Deformations of $ \Psi^{\rm t}(J=13\hbar,\theta=90^{\circ}) $,
are calculated to be $\beta=0.2661$ and $\gamma=0.015^{\circ}$.

We can consider whether this isomer is  yrast or not.
(Experimentally, 
the g-band is identified only up to $8\hbar$ \cite{TOI96}).
Xu et al. calculated that the non-collective prolate state
would be yrast at $J=13\hbar$ \cite{XWW00}.
Our self-consistent  calculations also show that 
$ \Psi^{\rm t}(J,\theta=90^{\circ}) $
 is lower in energy than principal axis rotation, 
 $ \Psi^{t}(J,\theta=0^{\circ}) $,
for $12\hbar \le J \le 34\hbar$,
so that in this region the high-$K$ isomeric states can be favoured
relative to  the collective rotation.

Let us use the term ``tilted rotation'' for the states with local
 minima, with $\theta \neq 0, 90^{\circ}$.
At $J=13\hbar$ (see Fig.3(b)), 
there is tilted rotation ($\theta\simeq 15^{\circ}$)
with an excitation energy of 2.593 MeV,
 which happens to be quite 
close to the experimental value of the $K^{\pi}=13^+$ isomer.
Numerical deformation values for this minimum are
$\beta=0.2704$ and $\gamma=4.62^{\circ}$, that is,
the shape is nearly prolate.
It is possible to consider the tilted rotation to
describe a rotational member of a high-$K$ state.
However, at this angular momentum, the potential energy curve
is shallow in the whole range of $\theta$ (see Fig.3(b)),
and the barrier height is the same order of magnitude
as uncertainties from the mean-field approximation.
The question as to whether the high-$K$  state should be
described as either a tilted rotation state  or 
a $z$-axis cranked state
$ \Psi^{t}(J,\theta=90^{\circ}) $ (or
coupling of them) 
should be answered by a quantum mechanical calculation  
by using angular momentum projection 
(and/or the generator coordinate method),
 which we plan to study in the future.
At this moment, the description of high-$K$ states
in the framework of self-consistent tilted-axis-cranking calculations
is reasonably good up to an accuracy of several hundred keV,
but more detailed studies are surely necessary.

For these shallow energy curves,
we can consider the wobbling motion to 
relate and cause transitions between high-$K$ isomers
and low-$K$ states such as the g-band.
For the energy curve at $J=13\hbar$,
there are three minima
at $\theta=0^{\circ}$ (principal axis rotation), 
$\theta=15^{\circ}$ (tilted rotation), and 
$\theta=90^{\circ}$ (possible high-$K$ states) within 600 keV.
If the first minimum represents a rotational member in the g-band 
and the third (and/or second) minimum represents the one 
in a high-$K$ band, 
then the corresponding band crossing is 
expected to show (experimentally) a strong coupling as evidence for
a realization of the wobbling motion.

In summary, 
we have performed  tilted-axis-cranked HFB calculations for $^{182}$Hf,
and investigated high-spin states near the yrast line up to $J=40\hbar$.
For our parameter set, 
the comparison of the experimental data for the g-band and a high-$K$
 isomeric state ($K^{\pi}=13^+$) with our calculation gives a reasonable
 agreement within the framework of the mean-field approximation.
With our modified single-particle model space 
based on  Kumar-Baranger's choice,
we found a new HFB solution with near-prolate deformation 
($\gamma\simeq -10^{\circ}$)  involving j$_{15/2}$ neutron alignment 
at high spin ($J \agt 17\hbar$).
We discussed the possibility 
of shape coexistence among  three states in principal axis rotation
with different gamma deformation:
 two with near-prolate shapes ($\gamma\simeq \pm 10^{\circ}$)
and the other with oblate shape ($\gamma\simeq 60^{\circ}$).
An analysis of the possible  backbends in yrare states
was also given.
In addition, we discussed the effect of wobbling motion as a coupling mode
between low-$K$ and high-$K$ states.

M.O. would like to acknowledge with thanks support 
from the Japan Society for the  Promotion of Science (JSPS). 
He also thanks  
Drs. N. Onishi, W. Nazarewicz, Y. R. Shimizu, and T. Nakatsukasa 
for useful discussions.
Part of the computations have been done using the vector parallel process type
super-computer, Fujitsu VPP700 at RIKEN and 
an Alpha workstation at Center for Nuclear Science (CNS), 
University of Tokyo.

\newpage
{\columnwidth\textwidth\csname@twocolumnfalse\endcsname
\begin{figure}[hbt]
\begin{center}
\leavevmode
\parbox{1.0\textwidth}
{\psfig{figure=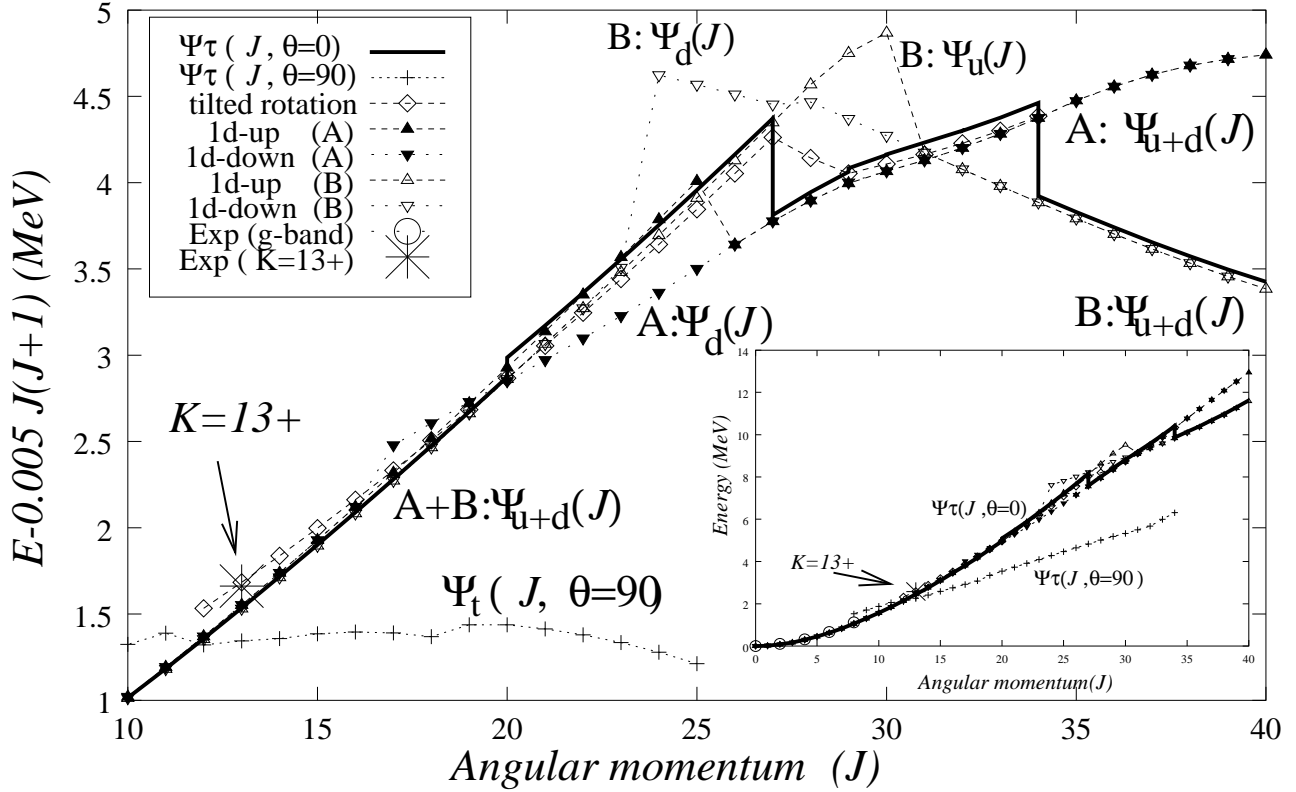,width=0.95\textwidth}}
\end{center}

\caption{
Energy spectrum near the yrast line obtained through the
tilted-axis-cranking calculations.
In the large graph, we show the energy with an arbitrary subtraction of 
rotational energy, $0.005 J(J+1)$ MeV for $10\hbar\le J\le 40\hbar$,
in order to see excited structures in detail.
In the small graph, the energy curve (without subtraction)
is shown for the full range of angular momentum $0\hbar \le J \le 40\hbar$. Experimental values for the g-band (open circle) 
and a $K^{\pi}=(13^+)$ isomer (asterisk) are plotted for a comparison.
One-dimensional up- and down-cranking are depicted by
upward and downward triangles, respectively.
When the up- and down-cranking results are the same,
the triangles are superposed to give the stars.
Two kinds of 1d-cranked calculations are shown as ``A'' and ``B'',
which are obtained by $\beta^{\rm ini}=0.270$ (open triangles)
 and $0.268$ (closed triangles), respectively.
``A+B'' means that the two 1d-cranking solutions ``A'' and ``B'' give 
 similar results. Also ``u+d'' means that the up- and down-cranking
calculations show no differences. 
The thick solid line indicates states of $\theta=0^{\circ}$ 
obtained in the 2d-up-cranking calculations 
($\Psi^{\rm t}(J,\theta=0^{\circ})$)
while plus signs denote 2d-up-cranking states 
$\Psi^{\rm t}(J,\theta=90^{\circ})$.
The tilted rotation minima are represented by diamonds
(see Fig.3(b)).
}
\end{figure}

\newpage
\begin{figure}[hbt]
\begin{center}
\leavevmode
\parbox{1.0\textwidth}
{\psfig{figure=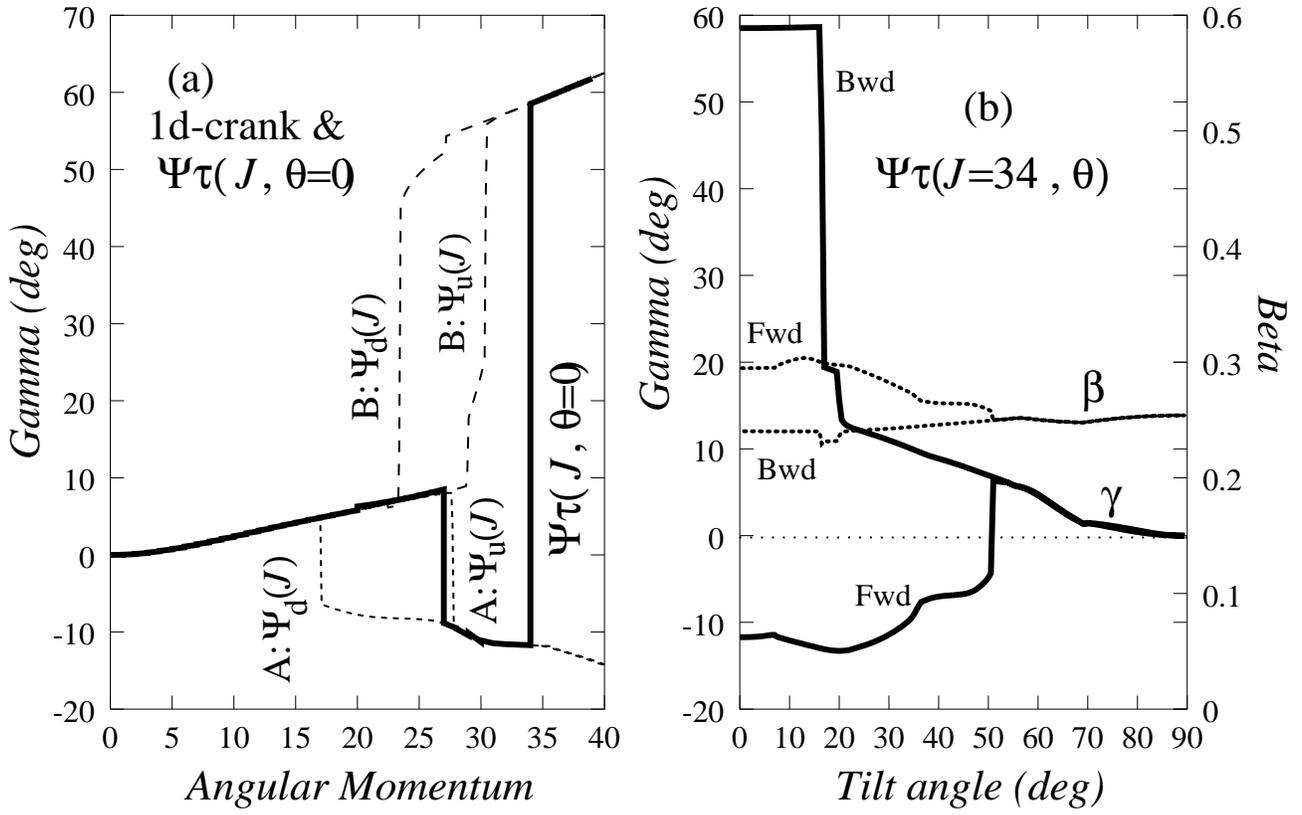,width=0.95\textwidth}}
\end{center}

\caption{
(a) Gamma deformation in 1d- and 2d-cranking calculations 
are shown with respect  to $J (= J_x)$. 
The symbols are the same as those in Fig.1.
(b) Gamma and beta values at $J=34\hbar$ 
in the 2d-cranking calculations
are shown with respect to $\theta$.
``Fwd'' means the forward tilting calculations,
while ``Bwd'' means  the backward calculations.
}
\end{figure}

\begin{figure}[hbt]
\begin{center}
\leavevmode
\parbox{0.9\textwidth}
{\psfig{figure=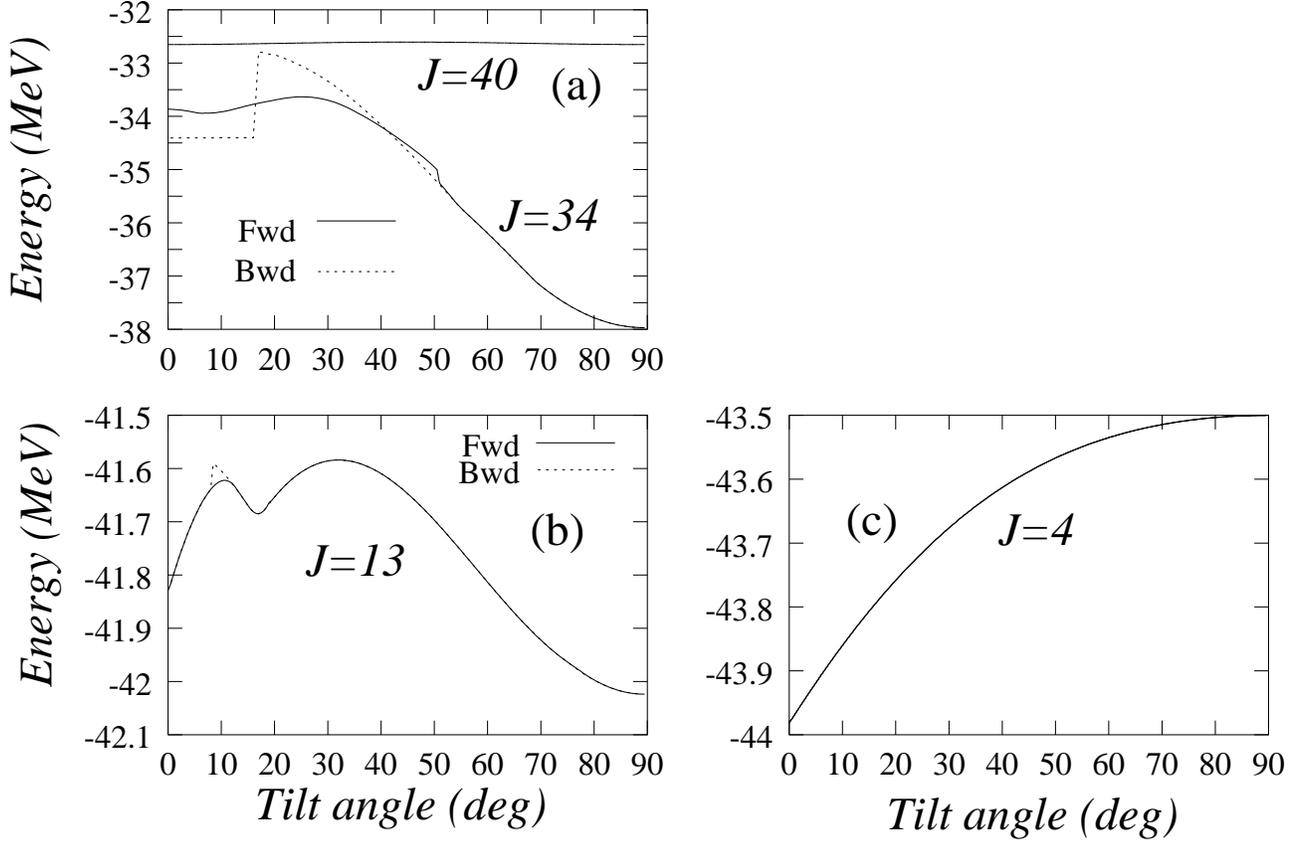,width=0.95\textwidth}}
\end{center}

\caption{Energy curves with respect to tilt angle $\theta$.
(a)energy curves at $J=34\hbar$ and $40\hbar$.
Three states having different gamma deformation are
energetically close  in the range  
$0^{\circ}\le \theta \le 40^{\circ}$.
For  $J=40\hbar$ 
there are two minima, at $\theta=0^{\circ}$ and $90^{\circ}$,
the barrier height between them being only about 50 keV.
(b)energy curve at $J=13\hbar$. 
For $12\hbar \le J \le 20\hbar$, there are three minima 
at $\theta=0^{\circ},90^{\circ}$, and $\simeq 15^{\circ}.$
The last minimum implies tilted rotation.
(c)energy curve at $J=4\hbar$. In the low-spin region ($J<12\hbar$),
there is only one minimum, at $\theta=0^{\circ}$, i.e.,
principal axis rotation.
}
\end{figure}
}
\end{document}